\documentclass[12pt]{article}
\usepackage{graphicx}

\begin{document}

\begin{center}
{\LARGE \bf Dashen-Frautschi Fiasco and \\[2ex]
Historical Roadmap for Strings}

\vspace{3ex}

Y. S. Kim\footnote{electronic address: yskim@physics.umd.edu}\\
Department of Physics, University of Maryland,\\
College Park, Maryland 20742, U.S.A.\\

\end{center}

\vspace{3ex}

\begin{abstract}

In 1964, Dashen and Frautschi published two papers in the Physical Review
claiming that they calculated the neutron-proton mass difference. It
was once regarded as a history-making calculation, in view of the fact that
the proton and neutron had been and still are regarded as the same particle
with different electromagnetic properties.  However, their calculation was
shown to be based on a bound-state wave function which violates the
localization condition in quantum mechanics.

There is one important lesson to be learned from the mistake made by
Dashen and Frautschi.  They did not pay much attention
to the fact that there are running waves and standing waves in quantum
mechanics.  The S-matrix formalism is based on running waves, while
bound-state problem are based on normalizable wave functions which are
standing waves.  Wave functions contained in the S matrix are analytic
continuations of running waves, and they do not in general satisfy the
localization condition for bound states.

These days, there are very serious questions raised against string theory.
Is string theory a form of quantum field theory?  Is it a physical theory
or only a mathematical exercise.  Is a new new Einstein going to emerge
from string theory?   It is pointed out that, if the distinction is
recognized between running and standing waves, string theory has its place
in the historical roadmap land-marked by Einstein, Heisenberg, Schr\"odinger,
Dirac, Wigner, and Feynman.

\end{abstract}

\newpage

\section{Introduction}\label{intro}

On April 29, at the 1965 spring meeting of the American Physical
Society in Washington, Freeman Dyson of the Institute of Advanced
Study (Princeton) presented an invited talk entitled ``Old and New
Fashions in Field Theory,'' and the content of his talk
was published in the June issue of the Physic Today~\cite{dyson65}.
This paper contains the following paragraph.

\begin{quote}
The first of these two achievements is the explanation
of the mass difference between neutron
and proton by Roger Dashen, working at the time as a graduate
student	under the supervision of Steve Frautschi.
The neutron-proton mass difference has for thirty years	been believed
to be electromagnetic in origin, and it offers a splendid
experimental test of any theory	which tries to cover the borderline
between electromagnetic and strong	interactions.	However,
no convincing theory of the mass-difference had appeared before
1964.  In this connection I exclude as unconvincing all	theories,
like the early	theory	of Feynman and Speisman, which use one
arbitrary cut-off parameter to	fit one	experimental number.
Dashen	for the first time made an honest calculation without arbitrary
parameters and got the right answer.  His method is a
beautiful marriage between old-fashioned electrodynamics and modern
bootstrap techniques.  He writes down the equations expressing
the fact that the neutron can be considered to be a bound state of a
proton with a negative pi meson, and the proton	a bound state of
a neutron with a positive pi meson, according to the bootstrap method.
Then into these equations he puts electromagnetic perturbations, the
interaction of a photon with both nucleon and pi meson, according to
the Feynman rules.  The calculation of the resulting mass difference is
neither long nor hard to understand, and in my opinion, it will become
a classic in the history of physics.
\end{quote}

Dyson was talking about the papers by R. F. Dashen and S. C. Frautschi
published in the Physical Review~\cite{df64}.  They use the S-matrix
formalism for bound states, and then derive a formula for a perturbed
energy level using the S-matrix quantities. Of course, they use
approximations because they are dealing with strong interactions.
There are however ``good'' approximations and ``bad'' approximations.

\begin{figure}[thb]
\centerline{\includegraphics[scale=0.5]{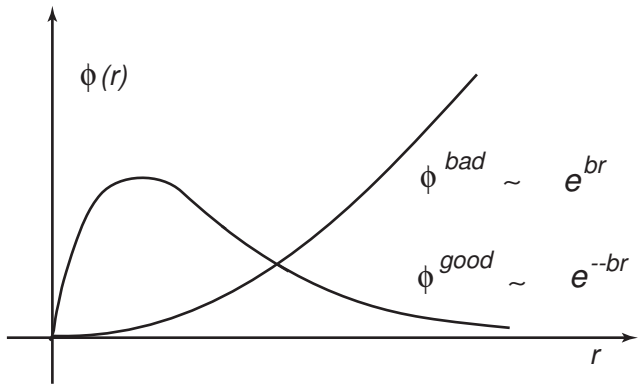}}
\vspace{5mm}
\caption{Good and bad wave functions contained in the S-matrix.
Bound-state wave functions satisfy the localization condition and are
good wave functions.  Analytic continuations of plane waves do not
satisfy the localization boundary condition, and become bad wave
functions at the bound-state
energy.}\label{goodbad}
\end{figure}

If we translate what they did into the language of the Schr\"odinger
picture of quantum mechanics, Dashen and Frautschi were using
the following approximation for the bound-state energy shift~\cite{kim66}
\begin{equation}\label{shift}
\delta E = \left(\phi^{good}, \delta V \phi^{bad} \right) ,
\end{equation}
where the good and bad bound-state wave functions are like
\begin{eqnarray}\label{wfs}
&{}& \phi^{good} \sim e^{-br} , \nonumber \\[2ex]
&{}& \phi^{bad} \sim e^{br} ,
\end{eqnarray}
for large values of $r$, as illustrated in Fig.~\ref{goodbad}.

The Schr\"odinger equation is a second-order differential equation
with two solutions.  If the energy positive, there are two running-wave
solutions.  For negative energies, the two solutions take ``good''
and ``bad" forms as indicated in Eq.(\ref{wfs}).  The good wave function
is normalizable and carries probability interpretation.  The bad wave
function is not normalizable, and cannot be given any physical
interpretation.  If we demand that this bad wave function disappear,
energy-levels become discrete.  This is how the bound-state energy
levels are quantized.

In the S-matrix formalism, the bound-states appear as poles in the complex
energy plane.  Those bound-state poles correspond to``good'' localized
wave functions in the Schr\"odinger picture.  At all other places, there
are unlocalized ``bad'' wave functions.  Dashen and Frautschi overlooked
this point when they used approximations in the S-matrix theory, and
ended up with the ``bad'' formula given in Eq.(\ref{shift}).

Yes!  Dashen and Frautschi made a serious mistake in 1964, but why do we
have to raise this issue now? The answer is very simple.  Since 1956,
there appeared many words in physics, including dispersion relations,
Mandelstam representation, Regge poles, S-matrix theory, N/D method,
and bootstrap dynamics.  Each of these words dominated the physics during
its own period.  However, they are now completely forgotten.  The reason
is that these temporary topics failed to place themselves to the historical
roadmap.

These days, there is one domineering word in physics, particularly in
particle theory.  It is of course {\it string theory}.  In view of the
recent history, the question arises whether this theory will also be lost
in history or will find its proper place in the roadmap landmarked by
Einstein, Heisenberg, Schr\"odinger, Dirac, Wigner, and Feynman.

The purpose of string theory is to understand the physics inside
relativistic particles.  Since particles are localized entities
in space-time, string theory necessarily has to deal with standing waves.
As in the case of of the Dashen-Frautschi fiasco, it is difficult to
produce standing waves within the S-matrix, or with Feynman diagrams.
For standing wave problems, it would be much easier to start with
standing waves.

We avoid standing waves because it is difficult to Lorentz-boost them,
while it is a trivial matter to write down plane waves in a
Lorentz-covariant manner. The expression for plane waves is
Lorentz-invariant.  Indeed, quantum field theory is possible because
the plane waves are invariant. Plane waves are running waves.

Standing waves are superposition of running waves in opposite
directions.  Do those superpositions remain invariant under Lorentz
boosts?  No, but they are still covariant.  This is the question
we wish to exploit from the 1971 paper of Feynman, Kislinger and
Ravndal~\cite{fkr71}.

In discussing standing waves, it is common to start with a hard-wall
potential, but mathematically it is more comfortable to use harmonic
oscillators.  Indeed, in their paper of 1971, Feynman {\it et al.}
start with a Lorentz-invariant harmonic oscillator equation~\cite{fkr71}.
This equation has many different solutions satisfying different boundary
conditions.  The solution they use is not normalizable in time-separation
variable, and cannot be given any physical interpretation.

On the other hand, it is possible to fix up their mathematics.  Their
Lorentz-invariant differential equation has  normalizable solutions
which can form a representation space for Wigner's little group for
massive particles~\cite{wig39,knp86}, whose transformations leave the
four-momentum of invariant.  The little group dictates the internal
space-time symmetry.

The normalizable oscillator solutions can be Lorentz-boosted and can
be used to show that quarks and partons are two different manifestations
of one covariant theory, as $E = p^{2}/2m$ and $E = cp$ are two different
limits of Einstein's energy-momentum relation.

These ingredients obtained from Feynman and Wigner could serve as the
starting point for standing waves in the Lorentz-covariant world.  If
string theory is going to survive in history, it should find its
own place in the historical raodmap.  If string theory is solve the
problem within a particle, it is necessarily a theory of standing
waves.  The issue is how to treat standing waves in Einstein's covariant
world.

In Sec.~\ref{lamb}, we point out what Dyson says is correct but he gives
a wrong example.  Indeed, quantum field theory can become effective when
combined with different branches of physics.  We illustrate this point
using the calculation of the Lamb shift.

In Sec~\ref{waves}, we note that there are running waves and standing
waves in quantum mechanics.  While it is trivial to Lorentz-boost
running waves, it requires covariance of boundary conditions to
understand fully standing waves.
In Sec.~\ref{wfsma}, we study the S-matrix constructed from non-relativistic
scattering theory.  The advantage of this approach is to trace the behavior of
wave functions in the S-matrix.  While the bound-state in the S-matrix theory
corresponds to a pole in the complex energy plane, it comes from the localization
of the wave function in the Schr\"odinger  picture.  Approximations in the
S-matrix does not always guarantee the localization of the bound-state wave
function.  This is the cause of the mistake Dashen and Frautschi made.

In Sec.~\ref{histobs}, we review the history of bound and scattering
states from comets and planets.  It was Newton who observed first that
the same physics is applicable to both comets and planets.  Again, the
same physics applies to scattering and bound states in nonrelativistic
quantum mechanics.  The question is what happens in Einstein's
Lorentz-covariant world.  Feynman diagrams work for scattering states,
but Feynman suggested the use of harmonic oscillators to approach
standing-wave problems in the Lorentz-covariant world.

In Sec~\ref{stsym}, we discuss the space-time symmetry applicable to
relativistic extended particles.  This symmetry is dictated by Wigner's
little group~\cite{wig39}.  We review the progress made in this field.
In Sec.~\ref{covham}, it is shown possible to construct the covariant
harmonic oscillator wave functions.  These wave functions can be
Lorentz-boosted, but they depend on the time-separation variable.  As
a physical application of this covariant harmonic oscillator formalism,

it is shown in Sec.~\ref{par} that the quark and parton models are two
different manifestation of the same covariant entity.  The most
controversial aspect of Feynman's parton picture is that the partons
interact incoherently with external signals.  This puzzle can be
explained within the framework of this oscillator formalism.

In Sec.~\ref{string},  in view of Feynman's efforts, we point out
that there is a well-defined place for string theory in the Einstein's
roadmap landmarked by Heisenberg, Schr\"odinger, Wigner, Dirac and
Feynman.  String theory belongs to bound states or standing waves in
the Lorentz-covariant world.

\section{Lamb Shift}\label{lamb}
It is a well-accepted view that Dyson gave a wrong example for the
combination of field theory and other branches of physics.  Yet, Dyson was
right in indicating that old-fashioned field theory could be more effective
when combined with other branches of physics which produce physics which
field theory cannot produce.

The Lamb-shift calculation is an excellent example.   Quantum electrodynamics
leads to a delta-function-like perturbing potential for the hydrogen atom.
The $s-state $ wave function does not vanish at the origin while the $p-state$
does.  Thus the first-order perturbation formula leads an energy shift for
the $s-state$ while leaving the $p-state$ unchanged.

Of course the delta-function perturbing potential is one of the triumphs of
quantum electrodynamics.  On the other hand, QED cannot produce localized
wave functions for the hydrogen atom.  We need the Schr\"odinger or Dirac
equation with the static coulomb potential to obtain those wave functions
needed for the Lamb-shift calculation.

In order to calculate those localized bound-state wave functions, we have
to impose boundary conditions at infinity.  This localization condition
is not properly addressed in QED or any other forms of quantum field theory.

Indeed, the wave functions used in the Lamb-shift calculation do not come
from quantum field theory.  Dyson was right in saying that field theory
could be more effective if coupled with other branches of physics.

\section{Running Waves and Standing Waves}\label{waves}
The Dashen-Frautschi fiasco teaches us an important lesson.  There are
running waves and standing waves in quantum mechanics.  Even though
the standing wave is a superposition of running waves, it requires
an additional care of boundary conditions.  We do not know how to
deal with this problem in the S-matrix formalism.

If not impossible, it is very difficult to formulate Lorentz boosts
for rigid bodies.  On the other hand, it seems to be feasible to
boost waves. Indeed, quantum mechanics allows us to look at extended
object as wave packets or standing waves.  Thus, we are interested
in boosting waves.  We should note here also that there are standing
and running waves.

\begin{figure}[thb]
\centerline{\includegraphics[scale=0.7]{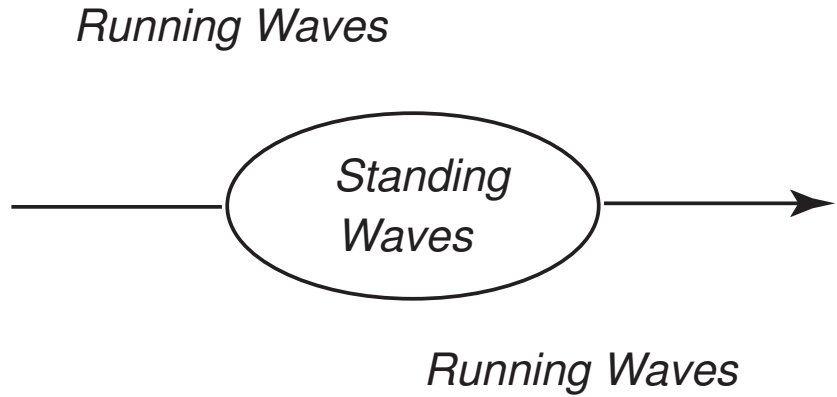}}
\vspace{5mm}
\caption{Running waves and standing waves in quantum theory.  It a
particle is allowed to travel from infinity to infinity, it corresponds
to a running wave according to the wave picture of quantum mechanics.
If, on the other hand, it is trapped in a localized region, we have
to use standing waves to interpret its location in terms of
probability distribution.}\label{dff11}
\end{figure}

Plane waves are running waves.  It is trivial to Lorentz-boost the
plane wave of the form
\begin{equation}
\exp{\left\{i\left(\vec{p}\cdot \vec{x} - p_{0} t \right)\right\}},
\end{equation}
because the exponent is invariant under Lorentz transformations.
However, what would happen when different waves are superposed?
Would the spectral function be covariant or invariant?
What would happen for standing waves which consist of superposition
of waves moving in opposite directions?

While quantum field theory based on Feynman diagrams starts with
running waves, quantum mechanics within a localized space-time region
deals with standing waves.  If string theory is set to solve the
problem inside particles, the physics of string theory is necessarily
the quantum mechanics of standing waves, as is indicated in
Fig.~\ref{dff11}.

In an attempt to obtain the answers to these questions, we can start
with some examples.  As usual in quantum mechanics, the first example
for standing waves should be a set of harmonic oscillator wave
functions.  With this point in mind, let us see what Feynman did for
harmonic oscillators in the relativistic regime.

\section{Wave Functions in S-matrix Theory}\label{wfsma}

Because of the success of quantum field theory which leads to the
S-matrix as the calculational tool, physicists were led to believe
all the problems in physics.  If this was not the case, the
S-matrix should be the source of information for all physical processes.

On the other hand, the S-matrix is derivable from two fundamental
theories.  One is of course quantum field theory.  The other is
the quantum mechanics of Schr\"odinger and Heisenberg, which leads to
non-relativistic potential scattering.

Indeed, much of the axioms or assumptions in the S-matrix theory is
based on analytic properties of the S matrix in the complex energy
plane.  In particular, the bound state in potential scattering
corresponds to a pole on in the negative energy axis.  Thus,
a perturbed energy-level corresponds to a displaced bound-state pole.
Since the S-matrix can be formulated in a Lorentz-covariant manner,
we can study the covariant picture of bound-states by studying the
S-matrix poles.

On the other hand, the S-matrix in both field theory and potential
scattering is basically a device to study scattering problems and
is therefore formulated in terms of running waves.  It does not
address the localization problem of bound-state wave functions.
In quantum field theory, the S-matrix cannot accommodate this
localization problem because there are no localized wave functions in
field theory.  In potential scattering we are dealing with solutions
of the Schr\"odinger equation which can accommodate both running and
standing waves.

Let us see how we can derive the first-order energy shift for perturbed
bound-state poles.  We start with an attractive potential $V{r}$.
For simplicity, we assume that it is spherically symmetric.  Then the
radial Schr\"odinger equation for the $\ell$th partial wave can be
written as
\begin{equation}\label{weq00}
\phi_{\ell}(r) = j_{\ell}(kr)+ {1 \over k}
\int_{0}^{r}  \left[n_{\ell}(kr)j_{\ell}(kr')-
j_{\ell} j(kr)n_{\ell}(kr')\right]
V_0(r')\phi_{\ell}(r')dr' ,
\end{equation}
where $\phi_{\ell}$ is the $\ell$th radial wave function. We
define $j_{\ell}(x)$ and $n_{\ell}(x)$ as
\begin{equation}
j_{\ell}= \left({\pi x \over 2}\right)^{1/2}J_{\ell + {1 \over 2}}(x),
\qquad
n_{\ell}= \left({\pi x \over 2}\right)^{1/2}N_{\ell + {1 \over 2}}(x),
\end{equation}
If we add a small perturbing potential $\delta V(r)$, then the
perturbed wave function $\psi(r)$ satisfies the equation
\begin{eqnarray}\label{weq11}
&{}&\hspace{-12mm} \psi_{\ell}(r) = j_{\ell}(kr)+ {1 \over k} \nonumber \\[2ex]
&{}& \hspace{3mm}\times \int_{0}^{r}  \left[n_{\ell}(kr)j_{\ell}(kr')-
j_{\ell} j(kr)n_{\ell}(kr')\right]
\left[V_0(r') + \delta V(r')\right]\psi_{\ell}(r')dr' ,
\end{eqnarray}
The perturbed wave function satisfies also the "full-Green's-function"
integral equation:
\begin{equation}\label{weq22}
\psi_{\ell}(r) = \phi_{\ell}(kr)+ {1 \over k}
\int_{0}^{r}  g_{\ell}(r,r') \delta V_0(r')\psi_{\ell}(r')dr' ,
\end{equation}
which, to lowest order in $\delta V(r)$, takes the form

\begin{equation}
\psi_{\ell}(r) = \phi_{\ell}(kr)+ {1 \over k}
\int_{0}^{r}  g_{\ell}(r,r') \delta V_0(r')\phi_{\ell}(r')dr' ,
\end{equation}
where $g_{\ell}(r',r)$ is the ``full Green's function'' constructed
from the solutions of Eq.(\ref{weq00}).

For simplicity, we now restrict ourselves to the S wave and drop the
subscript $\ell$. The generalization to higher partial waves
seems to be trivial. Then the integral equations in
Eqs. (\ref{weq00}), (\ref{weq11}), and (\ref{weq22}) take the following
forms.
\begin{eqnarray}
&{}& \phi (r) = j_{\ell}(kr)+ {1 \over k}
   \int_{0}^{r}  \sin k(r - r') V(r')\phi(r')dr' ,\nonumber \\[2ex]
&{}& \psi (r) = j_{\ell}(kr)+ {1 \over k}
\int_{0}^{r} \sin k(r - r') [V(r') + \delta V(r')]
\psi(r')dr' ,
\end{eqnarray}
and
\begin{equation}\label{weq77}
\psi (r) = \phi(r) + {1 \over k}
\int_{0}^{r} g(r,r')\delta V(r')\phi(r')dr' ,
\end{equation}
to lowest-order in $\delta V$.

In order to obtain the S matrix from the above solutions, we next
introduce the Jost functions $f(\pm k)$ and $F(\pm k)$, respectively,
for the perturbed and un-perturbed problems.
\begin{eqnarray}\label{jost00}
&{}& f(\pm k) = 1 + {1 \over k} \int_0^\infty
  [\exp{(\mp ikr)}]V_0(r) \phi(r) dr, \nonumber \\[2ex]
&{}& F(\pm k) = 1 + {1 \over k} \int_0^\infty
[\exp{(\mp ikr)}]\left[V_0(r) + \delta V(r)\right] \psi(r) dr.
\end{eqnarray}

In terms of these Jost functions we can now write the
phase shifts $\delta$ and $\eta$ for the original and perturbed
problems, respectively.
\begin{eqnarray}
\exp{\left(2i\delta \right)} = f(k)/ f(-k) ,\nonumber \\[2ex]
\exp{\left(2i\eta \right)} = F(k)/ F(-k) .
\end{eqnarray}

Therefor, in potential scattering, we study the S-matrix in terms
of the Jost functions $f(\pm k)$ or $F(\pm k)$.  In the S-matrix
theory, $f(k)$ is written as $N$ function or the numerator, and
$f(-k)$ as the D function or the denominator.  The S matrix is
therefore
\begin{equation}
S = N/D.
\end{equation}
Thus, the Jost-function approach in potential scattering corresponds
to the $N over D$ method in the S-matrix theory.

In the following discussions we will be led to study the
property of the Jost functions in the complex energy plane. We thus use $s$
as the energy variable, that is,
\begin{equation}
s=k^2 ,
\end{equation}
and adopt the following notation for the Jost functions.
\begin{eqnarray}
&{}& f^{(+)} (s) = f(k) , \qquad F^{(+)} (s) = F(k) ,  \nonumber \\[2ex]
&{}& f^{(-)} (s) = f(-k) , \qquad F^{(-)} (s) = F(-k) ,
\end{eqnarray}

Let us now assume that the unperturbed problem has
a bound state at $s = - s_0$, and therefore
\begin{equation}\label{pole1122}
f^{(-)}\left(-s_0 \right) = 0.
\end{equation}
For the perturbed system,
\begin{equation}\label{pole22}
F^{(-)}\left(- s_0- \delta s_0 \right) = 0,
\end{equation}
where $\delta s_0$ is the shift in the binding energy.
In order to ca1culate $\delta s_0$, we note from Eq.(\ref{jost00})
that $F^{(\mp)}(s)$ can be written as
\begin{equation}
F^{(\mp)}(s) = f^{(\mp)}(s) + \delta f^{(\mp)}(s),
\end{equation}
where
\begin{eqnarray}
&{}&\delta F^{(\mp)} = {1 \over k} \int_{0}^{\infty}[\exp{(\pm ikr)}]
   V_{0}[\psi(r) - \phi(r)] dr     \nonumber \\[2ex]
&{}& \hspace{10mm} + {1 \over k} \int_{0}^{\infty}[\exp{(\pm ikr)}]
 \delta V \phi(r) dr .
\end{eqnarray}

If the Born approximation is valid for the unperturbed
problem, both $\psi(r)$ and $\phi(r)$ become $\sin(kr)$,
and $\delta f^{(\mp)}(s)$ takes the following simple form:
\begin{equation}\label{pertu22}
\delta F^{(\mp)}(s) = {1 \over k} \int_{0}^{\infty}[\exp{(\pm ikr)}]
   \delta V(r) \sin(kr) dr .
\end{equation}
If, on the other hand, the Born approximation is not valid, we \
have to use Eq. (\ref{weq77}) for $[\psi(r) - \phi(r)]$, and the
above $\delta F^{(\mp)}(s)$ becomes
\begin{eqnarray}\label{pertu33}
&{}& \delta F^{(\pm)} = {1 \over k} \int_{0}^{\infty}[\exp{(\mp ikr)}]
   \delta V \phi(r)] dr     \nonumber \\[2ex]
&{}& +  {1 \over k^{2}} \int_{0}^{\infty}[\exp{(\mp ikr)}]
   V(r) dr \int_{0}^{\infty}g(r,r')  \delta V \phi(r) dr .
\end{eqnarray}
We return now to the bound-state condition of
Eq. (\ref{pole22}), which can be written as
\begin{equation}
f^{(-)}\left(- s_0 - \delta s_0\right) +
F^{(-)} \left(- s_0 - \delta s_0\right) = 0.
\end{equation}
By taking only the first-order terms in $\delta s_0$ and $\delta V(r)$,
we arrive at the following expression for $\delta s_0$.
\begin{equation}\label{pertu}
\delta s_0 = {\delta F^{(-)} \left(-s_0\right)
\over \left[f^{(-)}\right]'\left(-s_0\right) }
\end{equation}
This is the first-order correction to the binding energy.  This formula
is shown to be the same as $(\phi, \delta V \phi)$ for the square-well
potential where exact solutions are available for both scattering and
bound states~\cite{kim66}.

Dashen and Frautschi derived their bound-state formula from the $N/D$ method
which corresponds to the above formula in non-relativistic potential
scattering.  In so doing, they thought they could by-pass wave functions.
Furthermore, if we use the plane-wave approximation for the wave function
$\phi(r)$, $\delta F^{(\pm)}$ of Eq.(\ref{pertu33}) is reduced to the
Born-approximation formula of Eq.(\ref{pertu22}).  It is thus tempting to
use this approximation for the first try.  This is precisely what Dashen and
Frautschi did.

However, is the plane-wave approximation justified for bound-state problems?
The plane-wave solution in this case is $sin(kr)$ in Eq.(\ref{pertu22}).
When the energy becomes negative, the sine function becomes
\begin{equation}\label{sinkr}
\sin(kr) = {1 \over 2i}\left(e^{-br} - e^{br}\right) ,
\end{equation}
where $b = \sqrt{-s_{0}}$ .  For the bound state, $S_{0}$ is negative.
Thus, the above wave function increases exponentially for large values of r,
and is therefore a bad wave function.

What happens when we use the exact formula of Eq.(\ref{pertu33})?  The wave
function $\phi(r)$ is expected to be properly localized at the energy
$s = s_{0}$, and is drastically different from the bad wave function of
Eq.(\ref{sinkr}).

Dashen and Frautschi made this mistake because they did not realize that
there are standing waves and running waves in quantum mechanics.  For
present-day physics, let us see what lessons we can learn the historical
mistake made by Dashen and Frautschi.

\section{History of Scattering and Bound States}\label{histobs}

The issue of scattering versus bound states dates back to pre-Newton era.
It was not until Newton's second law and his law gravity that a unified
view of comets and planets was established, as is illustrated in
Table~\ref{histo}.

\begin{table}
\caption{History of Physics and Road Map for Strings.}\label{histo}
\begin{center}
\vspace{1mm}
\begin{tabular}{rccc}
\hline \\[-3.9mm]
\hline
{}\hspace{7mm} & {}  & {} &  {}\\
{}\hspace{7mm}  & {}& Unified & {} \\
{}\hspace{7mm} & Scattering & Physics & Bound States \\[4mm]\hline
{}\hspace{7mm} & {} & {} & {}\\

Before \hspace{7mm} & {} & {} & {} \\
Newton \hspace{7mm} & Comets & Unknown   & Planets \\[4mm]\hline
{}\hspace{7mm} & {} & {} & {}\\
Newton \hspace{7mm} & Hyperbola & Newton & Ellipse \\[4mm]\hline
{}\hspace{7mm} & {} & {} & {} \\
{}\hspace{7mm} & {} & {} & Quantized \\
Bohr \hspace{7mm} & Unknown & Unknown & Orbits \\[4mm]\hline
{}\hspace{7mm} & {} & {} & {}\\
Quantum \hspace{7mm} & Running  & Particle & Standing \\
Mechanics \hspace{7mm} & Waves & Waves  & Waves \\[4mm]\hline
{}\hspace{7mm} & {} & {} & {}\\
Feynman \hspace{7mm} & Diagrams & Unknown & Oscillators \\[4mm]\hline
{}\hspace{7mm} & {} & {} & {}\\
Future \hspace{7mm} & Running Waves  & One  & Standing Waves \\
Future \hspace{7mm} & * Fields  & Physics  & * Strings \\[2mm]
\hline
\hline
\end{tabular}
\end{center}
\end{table}

When Bohr established the law of orbit quantization of the hydrogen atom,
he was not able to explain scattering states.  Then, the Schr\"odinger
equation was able to generate scattering and bound-state solutions.
The Schr\"odinger picture of quantum mechanics accommodates the wave
nature of matter.  The Schr\"odinger equation is a second order
differential equation has two linearly independent solutions. For
scattering states, it is capable of two running waves in opposite
directions.  For bound states, one of the solutions vanish asymptotically
at infinite distance, while the other increases exponentially.  We then
demand that the wave function be normalizable.  In this way, we give
a localized probability distribution to the wave function. For scattering
states, we give an interpretation of probability current.

The present form of quantum mechanics was developed for non-relativistic
world.  Next step is to make it consistent with Einstein's special
relativity.  Since we are dealing with waves, we should learn how to
Lorentz-transform waves.  The question is how waves in one Lorentz frame
would look to observers in different frames.  The answer to this
question is trivial for plane waves of the form
\begin{equation}\label{planew}
\exp{(ip\dot x)} = \exp{\left\{i\left(\vec{p}\cdot \vec{x} - p_0 t\right)\right\}} .
\end {equation}
This form is invariant under Lorentz transformation, is usable to
observers in all different Lorentz frames.

It was possible before 1950 to develop quantum field theory because
field theory starts with plane waves of the form given
in Eq.(\ref{planew}).  For a superposition of plane waves:
\begin{equation}
\psi(x) = \int a(\vec{p}) e^{ip\cdot x} d^3 p ,
\end{equation}
The spectral function $a\left(\vec{p}\right)$ is defined as a covariant
quantity, but it has an additional physical interpretation through
second quantization or quantization of fields.  Yet,  the field theory
starts with plane waves, and this is the reason why it is possible to
develop a covariant scattering matrix.  It is also possible calculate
approximately the elements of this matrix using Feynman propagators
based on plane waves.  Feynman invented graphical approach to this
procedure, and the word "Feynman diagram" is very familiar to us.

The Feynman diagram is basically a physics of relativistic plane waves.
Plane waves are running waves.  How about standing waves?  Here we have
to take into account the fact that the standing wave consists of waves
moving in two opposite directions satisfying boundary conditions.
In quantum mechanics, we need those waves which can tell us about
localized probability distribution in different Lorentz frames. As in the
case of the hydrogen atom, the localization is defined in the
three space-like dimensions.  When the system is Lorentz-boosted those
the longitudinal component gets mixed with the time coordinate.  We
do not know how to deal with this problem.

\begin{figure}[thb]
\centerline{\includegraphics[scale=0.7]{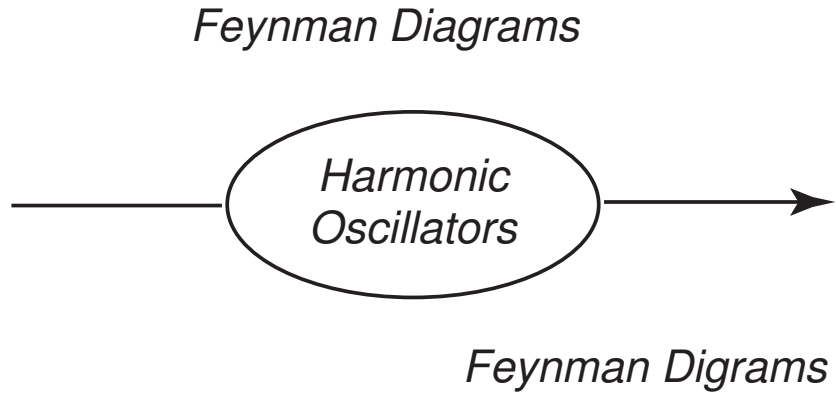}}
\vspace{5mm}
\caption{Feynman's roadmap for combining quantum mechanics with special
relativity.  Feynman diagrams work for running waves, and they provide
a satisfactory resolution for scattering states in Einstein's world.
For standing waves trapped inside an extended hadron, Feynman suggested
harmonic oscillators as the first step.}\label{dff33}
\end{figure}

It was clear that, from the talk Feynman gave in 1970~\cite{fey70}, that
feynman was aware of these problems.  It is quite common in physics that
physicists test their new theories by using harmonic oscillator, as in
the case of Einstein for specific heat, Heisenberg for quantum mechanics,
and Dirac for quantization of fields.  Feynman boldly suggested the use
of harmonic oscillator wave functions, instead of Feynman diagrams, to
approach bound-state problems in Einstein's relativistic world, as is
indicates in Fig~\ref{dff33}.  Feynman
talked about hadrons which are regarded as bound states of quarks.
With two of his students, Feynman published a paper in the Physical Review
D~\cite{fkr71} containing the content of his paper presented at the
Washington meeting.  This paper contains many mathematical inconsistencies,
which can be fixed up.

Feynman {\it et al.} start with a Lorentz-invariant differential equation
for the harmonic oscillator for the quarks bound together inside a hadron.
For the two-quark system, they write the wave function of the form
\begin{equation}
\exp{\left\{{-1 \over 2}\left(z^2 - t^2 \right) \right\}} ,
\end{equation}
where $z$ and $t$ are the longitudinal and time-like separations between the
quarks.  This form is invariant under the boost, but is not normalizable in
the $t$ variable.  Indeed, it is a ``bad'' wave function as in the case of
the Dashen-Frautschi fiasco.

On the other hand, the Gaussian form
\begin{equation}
\exp{\left\{{-1 \over 2}\left(z^2 + t^2 \right) \right\}}
\end{equation}
also satisfies Feynman's Lorentz-invariant differential equation.  This Gaussian
function is normalizable, but is not invariant under the boost.  However, the
word ``invariant'' is quite different from the word ``covariant.''  The above
form can be covariant under Lorentz transformations. We will get back to this
problem in Sec.~\ref{covham}.

Feynman {\it et al.} studied in detail the degeneracy of the three-dimensional
harmonic oscillators, and compared with the observed experimental data.  Their
work is complete and thorough.  However, they overlooked whether their oscillator
states are consistent with Wigner's little group which governs the internal
space-time symmetry of particles in Einstein's covariant regime.  They have not
reached this stage.

In Sec.~\ref{stsym}, we shall discuss see bound states or standing waves are
to be constructed, and the role of Wigner's little group for internal space-time
symmetries, before discussing the covariant harmonic oscillator formalism in
Sec.~\ref{covham}.

\section{Space-time Symmetries}\label{stsym}

In solving the problems in physics, we should decide what coordinate to use.
For spherical problems which are spherically symmetric, we should use
spherical but not Cartesian coordinate system.  For non-relativistic
problems we should use Galilean coordinate system.  For covariant
relativistic problem, we should use Lorentz-covariant system.
the problem.

Since Einstein introduced the Lorentz covariant space-time symmetry, his
energy momentum relation $E = \sqrt{p^2 + m^2}$ has been proven to be
valid for not only point particles, but also particles with internal
space-time structure, defined by quantum mechanics.  Particles can
have quantized spins if they are at rest of they are slowly moving.
If, on the other hand, the particle is massless and moves with speed of
light, it has its helicity which is the spin parallel to its momentum and
gauge degree of freedom.

\begin{table}[thb]
\caption{Massive and massless particles in one package.  Wigner's
little group unifies the internal space-time symmetries for massive and
massless particles.  It is a great challenge for us to find
another unification: the unification of the quark and parton pictures in
high-energy physics.}\label{einwig}
\vspace{3mm}
 \begin{center}
\begin{tabular}{lccc}
\hline
{}&{}&{}&{}\\
{} & Massive, Slow \hspace{6mm} & COVARIANCE \hspace{6mm}&
Massless, Fast \\[4mm]\hline
{}&{}&{}&{}\\
Energy- & {}  & Einstein's & {} \\
Momentum & $E = p^{2}/2m$ & $ E = [p^{2} + m^{2}]^{1/2}$ & $E = p$
\\[4mm]\hline
{}&{}&{}&{}\\
Internal & $S_{3}$ & {}  &  $S_{3}$ \\[-1mm]
Space-time &{} & Wigner's  & {} \\ [-1mm]
Symmetry & $S_{1}, S_{2}$ & Little Group & Gauge Trans. \\[4mm]\hline
{}&{}&{}&{}\\
Relativistic & {} & One  &  {} \\[-1mm]
Extended & Quark Model & Covariant  & Parton Model\\ [-1mm]
Particles & {} & Theory &{} {} \\[4mm]\hline

\end{tabular}

\end{center}
\end{table}

\begin{figure}[thb]
\centerline{\includegraphics[scale=0.7]{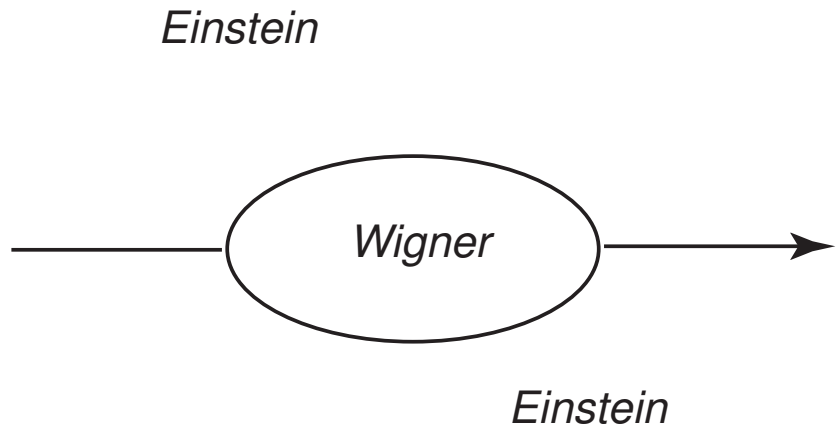}}
\vspace{5mm}
\caption{Wigner in Einstein's world.  Einstein formulates special
relativity whose energy-momentum relation is valid for point particles
as well as particles with internal space-time structure.  It was Wigner
who formulated the framework for internal space-time symmetries by introducing his
little groups whose transformations leave the four-momentum of a given particle
invariant.}\label{dff22}
\end{figure}

Table~\ref{einwig} summarizes the covariant picture of the present
particle world.  The second row of this table indicates that the spin
symmetry of slow particles and the helicity-gauge symmetry of massless
particles are two limiting cases of one covariant entity called Wigner's
little group.  This issue has been extensively discussed in the
literature~\cite{kiwi90jm}.

Let us then concentrate on the third row of Table~\ref{einwig}.  After
Einstein formulated his special relativity, a pressing problem was to see
whether his relativistic dynamics can be extended to rigid bodies as
in the case of Newton's sun and earth and their rotations.  As far as
we know, there are no satisfactory solutions to this problem.  However,
according to quantum mechanics, these extended objects are wave packets
or standing waves.  It might to easier to deal with waves in Einstein's
relativistic world.

Since Einstein worked with point particles when he was formulating his
special relativity and did not consider the physics inside the particles,
Some string theorists these days are calling for a new physics or a new
Einstein applicable to internal space-time symmetry and structure of
particles.

However, there also have been many respectable physicists in the past to
see whether Einstein prevails inside the particles.  As is illustrated in
Fig.~\ref{dff22}, Wigner formulated the concept of his little groups to
deal with the internal space-time symmetry of relativistic particles.

The development of the quark model for hadrons was another important step
toward understanding Einstein's covariance.  The proton is a quantum
bound state of quarks.  Since the proton these days can achieve a velocity
very close to that of light, and is a relativistic particle in the real
world.

While the proton is like a bound state when it is at rest, it appears
as a collection of partons when it moves with velocity close to that of
light.  As we shall discuss in Sec.~\ref{par}, partons have properties
which appear to be quite different from those of quarks.  Can we
produce a standing wave solution for the proton which can explain both
the quark model and the parton model?

\section{Can harmonic oscillators be made covariant?}\label{covham}

Quantum field theory has been quite successful in terms of
perturbation techniques in quantum electrodynamics.  However, this
formalism is based on the S matrix for scattering problems and useful
only for physical processes where a set of free particles becomes
another set of free particles after interaction.  Quantum field theory
does not address the question of localized probability distributions
and their covariance under Lorentz transformations.
The Schr\"odinger quantum mechanics of the hydrogen atom deals with
localized probability distribution.  Indeed, the localization condition
leads to the discrete energy spectrum.  Here, the uncertainty relation
is stated in terms of the spatial separation between the proton and
the electron.  If we believe in Lorentz covariance, there must also
be a time-separation between the two constituent particles.

\begin{figure}[thb]
\centerline{\includegraphics[scale=1.0]{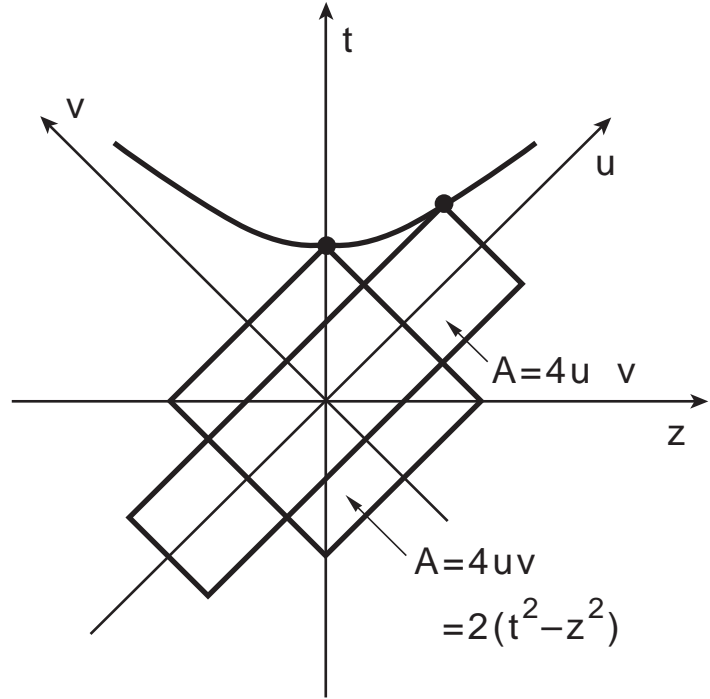}}
\vspace{5mm}
\caption{Lorentz boost in the light-cone coordinate
system.}\label{licone}
\end{figure}

Before 1964~\cite{gell64}, the hydrogen atom was used for
illustrating bound states.  These days, we use hadrons which are
bound states of quarks.  Let us use the simplest hadron consisting of
two quarks bound together with an attractive force, and consider their
space-time positions $x_{a}$ and $x_{b}$, and use the variables
\begin{equation}
X = (x_{a} + x_{b})/2 , \qquad x = (x_{a} - x_{b})/2\sqrt{2} .
\end{equation}
The four-vector $X$ specifies where the hadron is located in space and
time, while the variable $x$ measures the space-time separation
between the quarks.  According to Einstein, this space-time separation
contains a time-like component which actively participates as can be
seen from
\begin{equation}\label{boostm}
\pmatrix{z' \cr t'} = \pmatrix{\cosh \eta & \sinh \eta \cr
\sinh \eta & \cosh \eta } \pmatrix{z \cr t} ,
\end{equation}
when the hadron is boosted along the $z$ direction.
In terms of the light-cone variables defined as~\cite{dir49}
\begin{equation}
u = (z + t)/\sqrt{2} , \qquad v = (z - t)/\sqrt{2} ,
\end{equation}
the boost transformation of Eq.(\ref{boostm}) takes the form
\begin{equation}\label{lorensq}
u' = e^{\eta } u , \qquad v' = e^{-\eta } v .
\end{equation}
The $u$ variable becomes expanded while the $v$ variable becomes
contracted, as is illustrated in Fig.~\ref{licone}.

Does this time-separation variable exist when the hadron is at rest?
Yes, according to Einstein.  In the present form of quantum mechanics,
we pretend not to know anything about this variable.  Indeed, this
variable belongs to Feynman's rest of the universe.  In this report,
we shall see the role of this time-separation variable in the
decoherence mechanism.

\begin{figure}[thb]
\centerline{\includegraphics[scale=1.0]{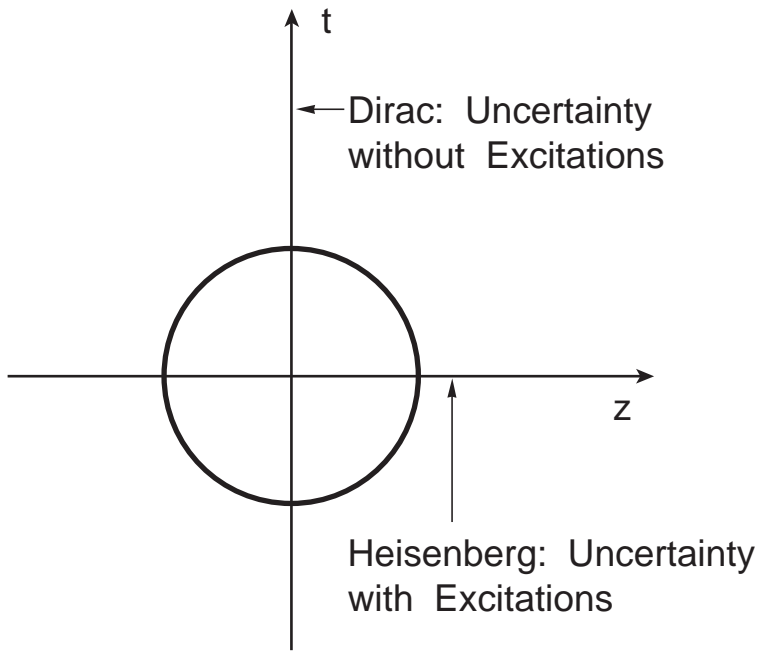}}
\vspace{5mm}
\caption{Space-time picture of quantum mechanics.  There
are quantum excitations along the space-like longitudinal direction, but
there are no excitations along the time-like direction.  The time-energy
relation is a c-number uncertainty relation.}\label{quantum}
\end{figure}

Also in the present form of quantum mechanics, there is an uncertainty
relation between the time and energy variables.  However, there are
no known time-like excitations.  Unlike Heisenberg's
uncertainty relation applicable to position and momentum, the time and
energy separation variables are c-numbers, and we are not allowed to
write down the commutation relation between them.  Indeed, the
time-energy uncertainty relation is a c-number uncertainty
relation~\cite{dir27}, as is illustrated in Fig.~\ref{quantum}

How does this space-time asymmetry fit into the world of
covariance~\cite{kn73}.  This question was studied in depth by the
present authors in the past.  The answer is that Wigner's $O(3)$-like
little group is not a Lorentz-invariant symmetry, but is a covariant
symmetry~\cite{wig39}.  It has been shown that the time-energy
uncertainty applicable to the time-separation variable fits perfectly
into the $O(3)$-like symmetry of massive relativistic
particles~\cite{knp86}.

The c-number time-energy uncertainty relation allows us to write down
a time distribution function without excitations~\cite{knp86}.
If we use Gaussian forms for both space and time distributions, we
can start with the expression
\begin{equation}\label{ground}
\left({1 \over \pi} \right)^{1/2}
\exp{\left\{-{1 \over 2}\left(z^{2} + t^{2}\right)\right\}}
\end{equation}
for the ground-state wave function.  What do Feynman {\it et al.}
say about this oscillator wave function?

\begin{figure}[thb]
\centerline{\includegraphics[scale=0.6]{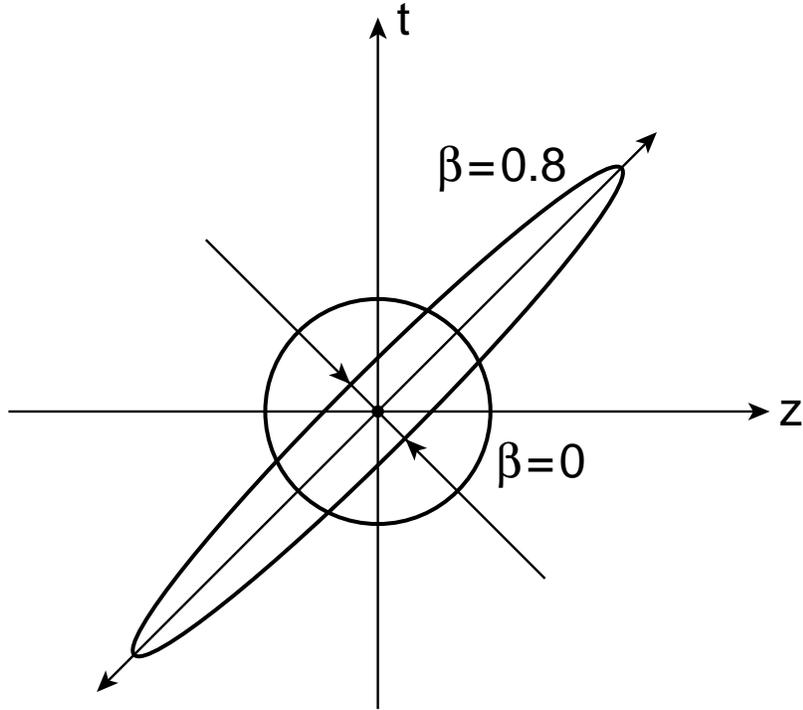}}
\caption{Effect of the Lorentz boost on the space-time
wave function.  The circular space-time distribution at the rest frame
becomes Lorentz-squeezed to become an elliptic
distribution.}\label{ellipse}
\end{figure}

In their classic 1971 paper~\cite{fkr71}, Feynman {\it et al.} start
with the following Lorentz-invariant differential equation.
\begin{equation}\label{osceq}
{1\over 2} \left\{x^{2}_{\mu} -
{\partial^{2} \over \partial x_{\mu }^{2}}
\right\} \psi(x) = \lambda \psi(x) .
\end{equation}
This partial differential equation has many different solutions
depending on the choice of separable variables and boundary conditions.
Feynman {\it et al.} insist on Lorentz-invariant solutions which are
not normalizable.  On the other hand, if we insist on normalization,
the ground-state wave function takes the form of Eq.(\ref{ground}).
It is then possible to construct a representation of the
Poincar\'e group from the solutions of the above differential
equation~\cite{knp86}.  If the system is boosted, the wave function
becomes
\begin{equation}\label{eta}
\psi_{\eta }(z,t) = \left({1 \over \pi }\right)^{1/2}
\exp\left\{-{1\over 2}\left(e^{-2\eta }u^{2} +
e^{2\eta}v^{2}\right)\right\} .
\end{equation}
This wave function becomes Eq.(\ref{ground}) if $\eta$ becomes zero.
The transition from Eq.(\ref{ground}) to Eq.(\ref{eta}) is a
squeeze transformation.  The wave function of Eq.(\ref{ground}) is
distributed within a circular region in the $u v$ plane, and thus
in the $z t$ plane.  On the other hand, the wave function of
Eq.(\ref{eta}) is distributed in an elliptic region with the light-cone
axes as the major and minor axes respectively.  If $\eta$ becomes very
large, the wave function becomes concentrated along one of the
light-cone axes.  Indeed, the form given in Eq.(\ref{eta}) is a
Lorentz-squeezed wave  function.  This squeeze mechanism is
illustrated in Fig.~\ref{ellipse}.

There are many different solutions of the Lorentz invariant differential
equation of Eq.(\ref{osceq}).  The solution given in Eq.(\ref{eta})
is not Lorentz invariant but is covariant.  It is normalizable
in the $t$ variable, as well as in the space-separation variable $z$.
How can we extract probability interpretation from this covariant
wave function?


\section{Feynman's Parton Picture}\label{par}

It is a widely accepted view that hadrons are quantum bound states
of quarks having localized probability distribution.  As in all
bound-state cases, this localization condition is responsible for
the existence of discrete mass spectra.  The most convincing evidence
for this bound-state picture is the hadronic mass spectra which are
observed in high-energy laboratories~\cite{fkr71,knp86}.

In 1969, Feynman observed that a fast-moving hadron can be regarded
as a collection of many ``partons'' whose properties appear to be
quite different from those of the quarks~\cite{fey69}.  For example,
the number of quarks inside a static proton is three, while the number
of partons in a rapidly moving proton appears to be infinite.  The
question then is how the proton looking like a bound state of quarks
to one observer can appear different to an observer in a different
Lorentz frame?  Feynman made the following systematic observations.

\begin{itemize}

\item[a.]  The picture is valid only for hadrons moving with
  velocity close to that of light.

\item[b.]  The interaction time between the quarks becomes dilated,
   and partons behave as free independent particles.

\item[c.]  The momentum distribution of partons becomes widespread as
   the hadron moves fast.

\item[d.]  The number of partons seems to be infinite or much larger
    than that of quarks.

\end{itemize}

\noindent Because the hadron is believed to be a bound state of two
or three quarks, each of the above phenomena appears as a paradox,
particularly b) and c) together.

In order to resolve this paradox, let us write down the
momentum-energy wave function corresponding to Eq.(\ref{eta}).
If we let the quarks have the four-momenta $p_{a}$ and $p_{b}$, it is
possible to construct two independent four-momentum
variables~\cite{fkr71}
\begin{equation}
P = p_{a} + p_{b} , \qquad q = \sqrt{2}(p_{a} - p_{b}) ,
\end{equation}
where $P$ is the total four-momentum.  It is thus the hadronic
four-momentum.

The variable $q$ measures the four-momentum separation between
the quarks.  Their light-cone variables are
\begin{equation}\label{conju}
q_{u} = (q_{0} - q_{z})/\sqrt{2} ,  \qquad
q_{v} = (q_{0} + q_{z})/\sqrt{2} .
\end{equation}
The resulting momentum-energy wave function is
\begin{equation}\label{phi}
\phi_{\eta }(q_{z},q_{0}) = \left({1 \over \pi }\right)^{1/2}
\exp\left\{-{1\over 2}\left(e^{-2\eta}q_{u}^{2} +
e^{2\eta}q_{v}^{2}\right)\right\} .
\end{equation}
Because we are using here the harmonic oscillator, the mathematical
form of the above momentum-energy wave function is identical to that
of the space-time wave function.  The Lorentz squeeze properties of
these wave functions are also the same.  This aspect of the squeeze
has been exhaustively discussed in the
literature~\cite{knp86,kn77par,kim89}.

When the hadron is at rest with $\eta = 0$, both wave functions
behave like those for the static bound state of quarks.  As $\eta$
increases, the wave functions become continuously squeezed until
they become concentrated along their respective positive
light-cone axes.  Let us look at the z-axis projection of the
space-time wave function.  Indeed, the width of the quark distribution
increases as the hadronic speed approaches that of the speed of
light.  The position of each quark appears widespread to the observer
in the laboratory frame, and the quarks appear like free particles.

\begin{figure}
\centerline{\includegraphics[scale=0.5]{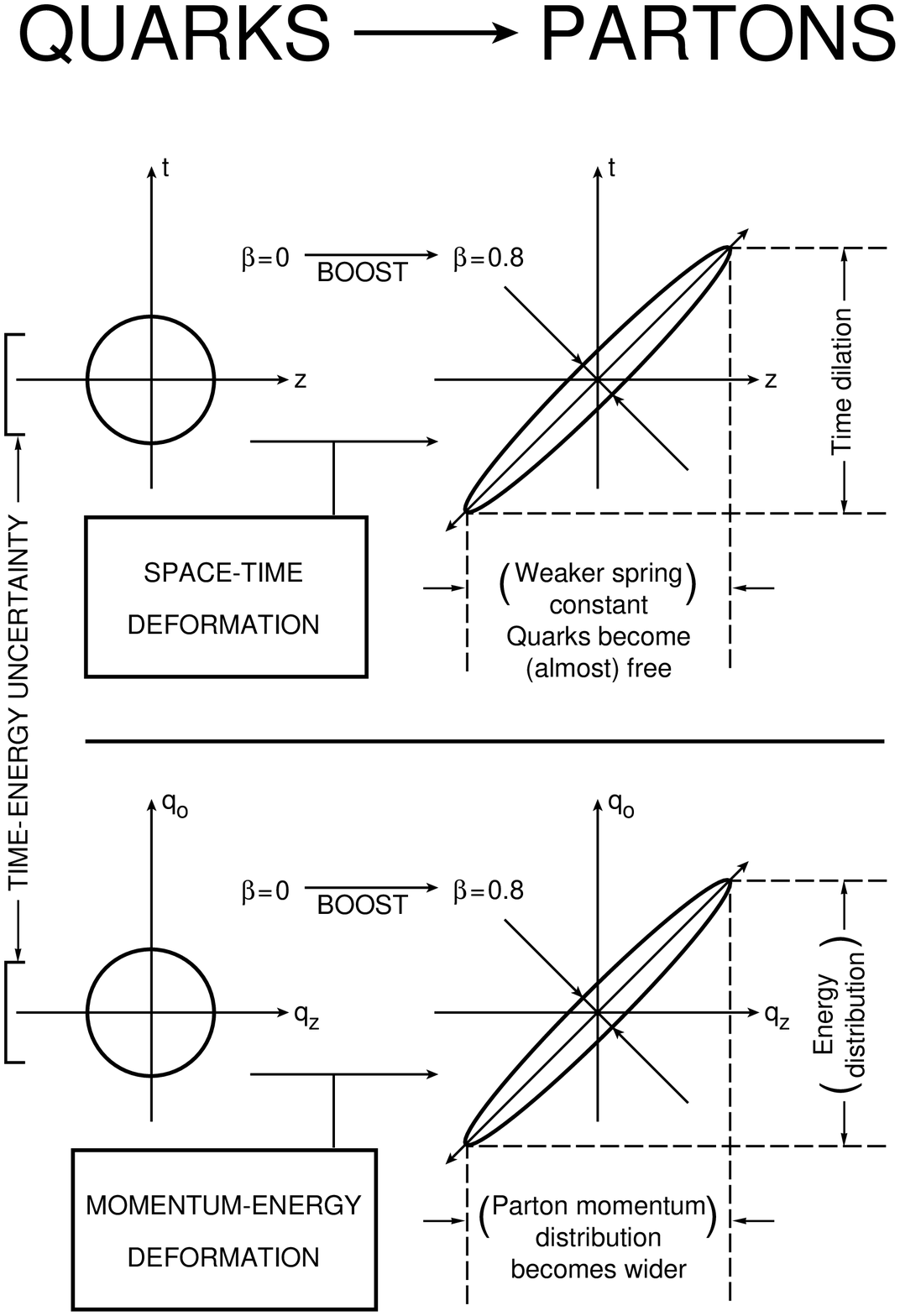}}
\vspace{5mm}
\caption{Lorentz-squeezed space-time and momentum-energy wave
functions.  As the hadron's speed approaches that of light, both
wave functions become concentrated along their respective positive
light-cone axes.  These light-cone concentrations lead to Feynman's
parton picture.}\label{parton}
\end{figure}

The momentum-energy wave function is just like the space-time wave
function, as is shown in Fig.~\ref{parton}.  The longitudinal momentum
distribution becomes wide-spread as the hadronic speed approaches the
velocity of light.  This is in contradiction with our expectation from
non-relativistic quantum mechanics that the width of the momentum
distribution is inversely proportional to that of the position wave
function.  Our expectation is that if the quarks are free, they must
have their sharply defined momenta, not a wide-spread distribution.

However, according to our Lorentz-squeezed space-time and
momentum-energy wave functions, the space-time width and the
momentum-energy width increase in the same direction as the hadron
is boosted.  This is of course an effect of Lorentz covariance.
This indeed is the key to the resolution of the quark-parton
paradox~\cite{knp86,kn77par}.

\begin{figure}
\centerline{\includegraphics[scale=0.6]{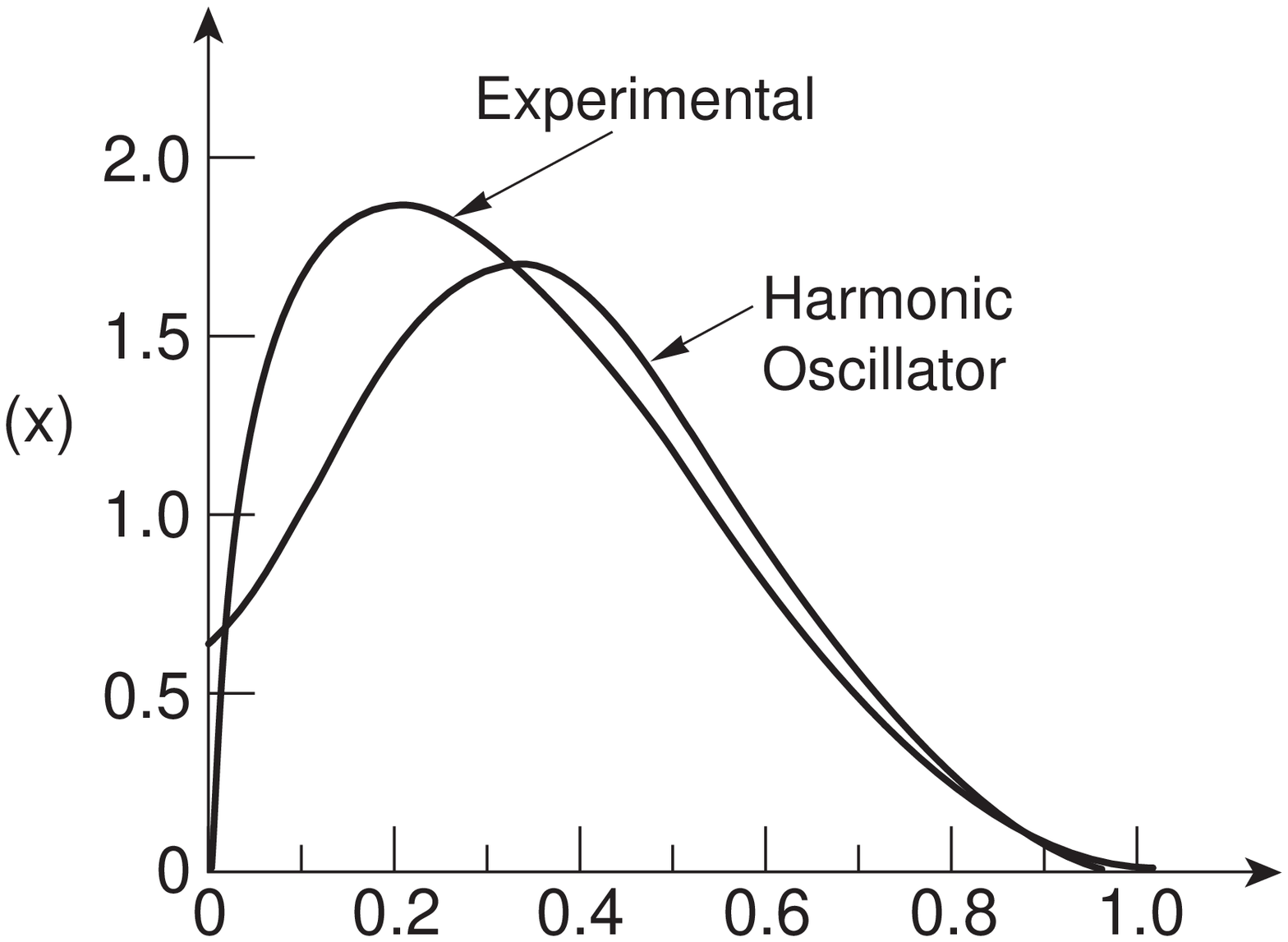}}
\vspace{5mm}
\caption{Parton distribution function.
Theory and experiment.}\label{hussar}
\end{figure}

After these qualitative arguments, we are interested in whether
Lorentz-boosted bound-state wave functions in the hadronic rest
frame could lead to parton distribution functions.  If we start with
the ground-state Gaussian wave function for the three-quark wave
function for the proton, the parton distribution function appears
as Gaussian as is indicated in Fig.~\ref{hussar}.  This Gaussian  form
is compared with experimental distribution also in Fig.~\ref{hussar}.

For large $x$ region, the agreement is excellent, but the agreement is
not satisfactory for small values of $x$.  In this region, there is
a complication called the ``sea quarks.''  However, good sea-quark physics
starts from good valence-quark physics.  Figure~\ref{hussar} indicates
that the boosted ground-state wave function provides a good valence-quark
physics.

\section{Historical Destiny for Strings}\label{string}

Together with Marilyn Noz, the present author has been interested
in question of covariant harmonic oscillators since 1973~\cite{kn73}.
We started with the covariant oscillator wave function as a purely
phenomenological mathematical instrument.  We then noticed that
the covariant oscillator formalism can serve as a representation
of the Wigner's little group for massive particles, capable of
the fundamental symmetry representation for relativistic particles.
This allows us to deal with the c-number time-energy uncertainty
relation without excitations.  Furthermore, the Lorentz-boosted
Gaussian wave function produces a parton distribution in satisfactory
agreement with experimental data.

\begin{figure}[thb]
\centerline{\includegraphics[scale=0.7]{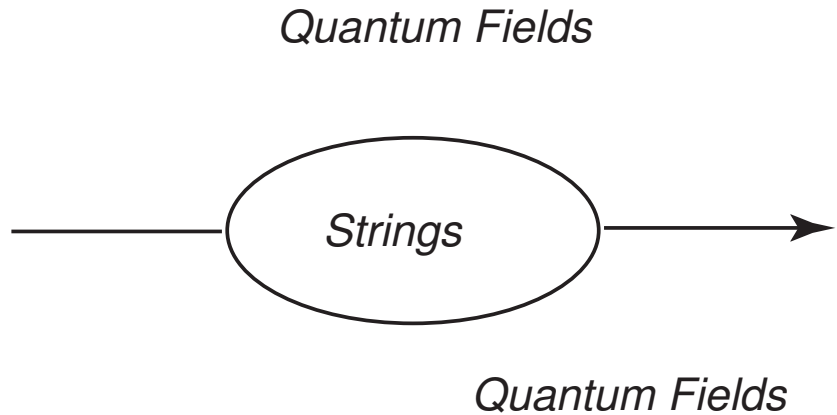}}
\vspace{5mm}
\caption{Roadmap for strings.  If quantum fields are running waves,
strings constitute the language for standing waves insider a particle
localized in finite space-time region.  This is the most appropriate place
in Einstein's roadmap landmarked by Wigner and Feynman.}\label{dff44}
\end{figure}

What are then Feynman's contributions to this subject?  In addition
to the formulation of the parton picture, he suggested the use of
harmonic oscillator wave functions to understand bound-state problems
in the covariant regime.   Then where does the Feynman diagram stand
in his scheme?  Feynman diagrams start with plane waves which are
running waves.  Harmonic-oscillator wave functions are standing waves.
For standing waves, we have to take care of the covariance of boundary
conditions or spectral functions.  This is precisely what we are
reporting in this report.

It is gratifying to note that there is only one covariant quantum
mechanics for both scattering and bound states.  In both cases, we
deal with waves in the covariant world.  For scattering
states, we are dealing with asymptotically free waves and Feynman
diagrams.  For bound states, we should start with standing waves.
The covariant harmonic oscillator wave functions could constitute a
complete set of wave functions we can start with.

It is our understanding that the purpose of string theory is to
understand the physics inside particles, as is indicated in
Fig.~\ref{dff44}. Since particles are localized entities in
the space-time region, string theory is  necessarily a physics of
standing waves if we are to preserve the present form of quantum
mechanics.  The Lorentz covariance of the standing waves should
be the major issue in string theory.

\end{document}